\documentclass{iopart}

\usepackage{graphicx}

\newcommand{\hide}[1]{}

\begin{document}

\title[PET monitoring with $^3$He beams]{PET monitoring of cancer therapy with 
$^3$He and $^{12}$C beams: a study with the GEANT4 toolkit}
    
\author{Igor~Pshenichnov$^{1,2}$, Alexei~Larionov$^{1,3}$, Igor~Mishustin$^{1,3}$ 
and Walter~Greiner$^1$}  
\address{$^1$ Frankfurt Institute for Advanced Studies, Johann Wolfgang Goethe University, 
60438 Frankfurt am Main, Germany}
\address{$^2$ Institute for Nuclear Research, Russian Academy of Science, 117312 Moscow, Russia}
\address{$^3$ Kurchatov Institute, Russian Research Center, 123182 Moscow, Russia}

\begin{abstract}
We study the spatial distributions of $\beta^+$-activity produced by therapeutic beams of $^3$He 
and $^{12}$C ions in various tissue-like materials. The calculations were performed within a  
Monte Carlo model for Heavy-Ion Therapy (MCHIT) based on the GEANT4 toolkit.  
The contributions from positron-emitting nuclei with $T_{1/2}>10$~s, 
namely $^{10,11}$C, $^{13}$N, $^{14,15}$O, $^{17,18}$F and $^{30}$P, were 
calculated and compared with experimental data obtained during and after irradiation, 
where available. Positron emitting nuclei are created by $^{12}$C beam in 
fragmentation reactions of projectile and target nuclei.
This leads to a $\beta^+$-activity profile characterised by a noticeable 
peak located close to the Bragg peak in the corresponding depth-dose distribution. 
This can be used for dose monitoring in carbon-ion therapy of cancer.
On the contrary, as the most of positron-emitting nuclei are produced by $^3$He beam in target fragmentation
reactions, the calculated  total  $\beta^+$-activity during or soon after the irradiation period 
is evenly distributed within the projectile range. However, we predict also the presence of
$^{13}$N, $^{14}$O, $^{17,18}$F created in charge-transfer reactions by low-energy $^3$He ions
close to the end of their range in several tissue-like media. The time evolution of 
$\beta^+$-activity profiles was investigated for both kinds of beams.
We found that due to the production of $^{18}$F nuclide the $\beta^+$-activity profile 
measured 2 or 3 hours after irradiation with $^{3}$He ions will have a distinct peak correlated with
the maximum of depth-dose distribution. We also found certain advantages of low-energy $^{3}$He 
beams over low-energy proton beams for reliable PET monitoring during
particle therapy of shallow located tumours. In this case the distal edge of 
$\beta^+$-activity distribution from $^{17}$F nuclei clearly marks the range of 
$^{3}$He in tissues.  
\end{abstract}

\submitto{\PMB}
% Key words: theory and algorithms in physics of heavy-ion therapy, Monte Carlo applications, simulation

\ead{pshenich@fias.uni-frankfurt.de}

% ( )\hide{~\cite{  }}

\section{Introduction}

Beams of protons and carbon ions are used in particle therapy of deep-seated tumours 
for conformal irradiation of a tumour volume while sparing surrounding
healthy tissues and organs at risk (Castro \etal 2004,\hide{~\cite{Castro:etal:2004}} 
Amaldi and Kraft 2005\hide{~\cite{Amaldi:and:Kraft:2005}}). 
New facilities for proton and ion therapy of cancer are planned or under 
construction in France (Bajard \etal 2004)\hide{~\cite{Bajard:etal:2004}},
in Italy  (Amaldi 2004)\hide{~\cite{Amaldi:2004}}, 
in Austria (Griesmayer and Auberger 2004)\hide{~\cite{Griesmayer:and:Auberger:2004}} and 
in Germany (Haberer \etal 2004)\hide{~\cite{Haberer:etal:2004}}. Along with other facilities 
the Heidelberg Ion Therapy Center (HIT) (Haberer \etal 2004,\hide{~\cite{Haberer:etal:2004}}
Heeg \etal 2004\hide{~\cite{Heeg:etal:2004}})
will use several beams for treatment: protons, $^3$He, carbon and oxygen nuclei. 
There oncologists will have 
a variety of treatment tools at their disposal, while each treatment option
will be characterised by its specific effectiveness, possible side effects and treatment costs. 
Appropriate quality assurance methods should be also developed specifically for each kind of treatment. 

In particular, the Positron Emission Tomography (PET) during or after irradiation provides the  
possibility to monitor the delivered dose. The PET monitoring methods in proton and ion therapy can be
divided into two categories: (i) based on tracing the positron emitting nuclei,
e.g. $^{10}$C, $^{11}$C and $^{15}$O,  created by proton beams in tissues due to 
fragmentation of {\it target nuclei}; (ii) based on tracing the positron emitting nuclei, 
$^{10}$C and $^{11}$C, created in fragmentation reactions of $^{12}$C {\it beam nuclei.} 
The spatial distribution of positron emitting nuclei is measured by 
detecting gamma pairs from  $e^+ e^- \to \gamma\gamma$ annihilation events. 
By comparing the measured $\beta^+$-activity 
distribution with the distribution calculated for the planned dose, one can control 
the accuracy of the actual treatment.

Following extensive theoretical and experimental studies with carbon beams
(Enghardt \etal 1992,\hide{~\cite{Enghardt:etal:1992}} Pawelke \etal 1996,\hide{~\cite{Pawelke:etal:1996}} 
Pawelke \etal 1997,\hide{~\cite{Pawelke:etal:1997}} 
P\"onisch \etal 2004,\hide{~\cite{Poenisch:etal:2004}}
Parodi 2004\hide{~\cite{Parodi:2004}}), in-beam PET monitoring is successfully used 
in carbon-ion therapy at GSI, 
Darmstadt, Germany (Enghardt \etal 2004,\hide{~\cite{Enghardt:etal:2004}} 
Schulz-Ertner \etal 2004\hide{~\cite{Schulz-Ertner:etal:2004}}).
Similar approaches can be used for monitoring of proton therapy, as shown early by 
Bennett \etal 1975,\hide{~\cite{Bennett:etal:1975}} Bennett \etal 1978\hide{~\cite{Bennett:etal:1978}}
and later by Oelfke \etal 1996,\hide{~\cite{Oelfke:etal:1996}}   
Parodi and Enghardt 2000,\hide{~\cite{Parodi:and:Enghardt:2000}}  
Parodi \etal 2002\hide{~\cite{Parodi:etal:2002}}, Parodi 2004\hide{~\cite{Parodi:2004}},
Nishio \etal 2005\hide{~\cite{Nishio:etal:2005}}, 
Parodi \etal 2007a\hide{~\cite{Parodi:etal:2007a}} 
and Parodi \etal 2007b.\hide{~\cite{Parodi:etal:2007b}} 
PET images from proton and carbon-ion therapy were also studied in experiments by 
Hishikawa \etal 2004\hide{~\cite{Hishikawa:etal:2004}} at Hyogo ion 
therapy centre in Japan (Hishikawa \etal 2002\hide{~\cite{Hishikawa:etal:2002}}).

Beams of nuclei lighter than carbon, e.g. $^{3}$He, $^{4}$He or $^{7}$Li, are also of clinical
interest. This is shown, in particular, by Furusawa \etal 2000\hide{~\cite{Furusawa:etal:2000}} and 
Kempe \etal 2007\hide{~\cite{Kempe:etal:2007}}. 
An advantage of $^{3}$He nuclei consists in their specific $Z/A=2/3$ ratio which helps to 
protect the $^3$He beam from contamination with $^4$He, $^{12}$C, $^{16}$O nuclei. 
The feasibility of in-beam PET for $^3$He therapy was
demonstrated for the first time in experiments by Fiedler \etal 2006.\hide{~\cite{Fiedler:etal:2006}} 

In the present work we use a Monte Carlo model for Heavy-Ion Therapy 
(MCHIT) (Pshenichnov \etal 2005, 2006)\hide{~\cite{Pshenichnov:etal:2005,Pshenichnov:etal:2006}} based
on the GEANT4 simulation toolkit (Agostinelli \etal 2003, 
Allison \etal 2006)\hide{~\cite{Agostinelli:etal:2003,Allison:etal:2006}}
to study the $\beta^+$-activity profiles induced by $^3$He and $^{12}$C  beams in tissue-like media.
We argue that specific nuclear reactions, namely {\it proton pick-up by target nuclei,} 
play certain role in production of positron emitting nuclei by $^3$He beams in 
addition to previously studied nuclear fragmentation reactions.  
In Section~\ref{model} we describe the physical models from the GEANT4 toolkit used to 
build MCHIT. In Section~\ref{time} the time-dependent analysis of the 
$\beta^+$-activity distributions induced by $^3$He and $^{12}$C  beams in graphite, water and PMMA
phantoms is presented. In Section~\ref{comparison} 
calculational results are compared with experimental data obtained
by Fiedler \etal 2006\hide{~\cite{Fiedler:etal:2006}}. 
Results for homogeneous phantoms with stoichiometric composition of muscle and bone
tissues are presented in Section~\ref{tissues}. 
The calculated distributions of $\beta^+$-activity induced by low energy proton, 
$^3$He and $^{12}$C beams are discussed in Section~\ref{LowEnergy} with emphasis on the role
of proton pick-up reactions induced by $^3$He. In Section~\ref{discussion} the reliability of
MCHIT results is verified by comparison with available experimental data on 
specific reaction cross sections and isotope yields in thick targets. 
Section~\ref{summary} contains summary and conclusions.

\section{GEANT4 physics models used in MCHIT}\label{model}

We have used the version 8.2  of the GEANT4 toolkit (GEANT4-Webpage 2006)\hide{~\cite{GEANT4-Webpage:2006}} 
to build a Monte Carlo model for Heavy-Ion Therapy (MCHIT). The model is intended for 
calculating the spatial distributions of dose and $\beta^+$-activity from beams of light
nuclei (from protons to oxygen ions) in homogeneous tissue-like media.
The phantom material and size, as well as beam parameters such as energy spread,
transverse beam size, emittance, angular divergence, can be set via user interface commands. 

In MCHIT the energy loss and straggling of primary and secondary charged particles due to
interaction with atomic electrons  is described via a set of models 
called 'standard electromagnetic physics'. Multiple scattering due to electromagnetic
interactions with atomic nuclei is also included in simulations.
 
In each simulation step, the ionisation energy loss of a charged particle is calculated according to the
Bethe-Bloch formula. The average excitation energy of the water molecule was set to 77 eV, 
i.e. to the value which better describes the set of available data on 
depth-dose distributions for therapeutic proton and carbon-ion beams. This parameter
was taken 68.5 eV for PMMA, 78 eV for graphite,  86.5 eV for bone tissue and 70.9 eV for muscle tissue. 

Two kinds of hadronic interactions are considered in the MCHIT model:
(a) elastic scattering of hadrons on target protons and nuclei, which dominate 
at low projectile energies, and (b) inelastic nuclear reactions induced by 
fast hadrons and nuclei (GEANT4-Documents 2006)\hide{~\cite{GEANT4-Documents:2006}}.

The overall probability of hadronic interactions for nucleons and nuclei propagating 
through the medium depends on the total inelastic cross section 
for proton-nucleus and nucleus-nucleus collisions.
Parametrised equations by Wellisch and Axen 1996\hide{~\cite{Wellisch:and:Axen:1996}} 
that best fit experimental data were used  to describe the total reaction
cross sections in nucleon-nucleus collisions. 
Systematics by Tripathi \etal 1997\hide{~\cite{Tripathi:etal:1997}} 
and Shen \etal 1989\hide{~\cite{Shen:etal:1989}} 
for the total nucleus-nucleus cross sections were used for calculating the probability of 
nucleus-nucleus collisions.

In the MCHIT model the inelastic interaction of nucleons below 20 MeV 
is simulated by means of data driven models. Above 20 MeV the exciton-based precompound
model is invoked (Agostinelli \etal 2003, 
Allison \etal 2006)\hide{~\cite{Agostinelli:etal:2003,Allison:etal:2006}}.
For hadrons and nuclei with the energies above 80 MeV, we employed
the binary cascade model  (Folger \etal 2004)\hide{~\cite{Folger:etal:2004}}.
Exited nuclear remnants are created after the first cascade stage of interaction. Therefore, appropriate
models for describing the de-excitation process have to be involved into simulation.   
The Weisskopf-Ewing model (Weisskopf and Ewing 1940)\hide{~\cite{Weisskopf:and:Ewing:1940}} 
was used to describe the evaporation of nucleons from residual nuclei at relatively 
low excitation energies, below 3 MeV per nucleon. 
The Statistical Multifragmentation Model (SMM) by
Bondorf \etal 1995\hide{~\cite{Bondorf:etal:1995}}
was used at excitation energies above 3 MeV per nucleon
to describe the multi-fragment break-up of highly-excited residual nuclei.
The SMM includes as its part the Fermi break-up model, which  
describes the explosive decay of highly-excited light nuclei.

\section{Time-dependent analysis of the $\beta^+$ activity}\label{time}

We follow the approach of Parodi \etal 2002\hide{~\cite{Parodi:etal:2002}}
in calculating the time dependence of $\beta^+$-activity induced by therapeutic beams.
As reported by Fiedler \etal 2006\hide{~\cite{Fiedler:etal:2006}}, the time structure of 
ion beams provided by the GSI synchrotron consists of repeated particle extractions (spills) and pauses.
It is assumed in calculations, that each beam spill has duration of $\tau_s$ with
the average intensity of $J$ (ions/s) during the beam extraction.
The irradiation procedure consists of $N$ spills with pauses between subsequent 
spills of $\tau_p$, as given in Table~\ref{tab:irradiation_parameters}.
Both $\tau_s$ and $\tau_p$ are in the range of 1-3~s.

\begin{table}[htb]
\caption{\label{tab:irradiation_parameters} Beam parameters for  
207.92 A MeV $^3$He and 337.5 A MeV  $^{12}$C used for irradiation of
graphite, water and PMMA phantoms by Fiedler {\it et al} 2006\hide{~\cite{Fiedler:etal:2006}}.
It is denoted: $N$ --- number of spills , $J$ --- beam intensity during each spill, 
$\tau_s$ --- spill duration, $\tau_p$ --- duration of a pause between spills. }
\begin{indented}
\item[]\begin{tabular}{@{}lllllll}
\br
        & Projectile & Phantom   &  $N$   &  $J$               &  $\tau_s$ & $\tau_p$ \\ 
        &            & material   &        &($10^8$ s$^{-1}$) &  (s)      & (s)            \\
\mr
        & $^3$He     & graphite & 120   & 1.9              &  1.39     & 3.10           \\
        & $^3$He     & water    & 99    & 2.0              &  1.35     & 3.14           \\
        & $^3$He     & PMMA     & 120   & 2.0              &  1.37     & 3.12           \\
        & $^{12}$C   & graphite & 120   & 0.9              &  2.20     & 2.29           \\
        & $^{12}$C   & water    & 120   & 0.9              &  2.20     & 2.29           \\
        & $^{12}$C   & PMMA     & 120   & 0.9              &  2.19     & 2.30           \\ 
\br
\end{tabular}
\end{indented}
\end{table}

The depth distributions $dn_i(z)/dz \equiv f_i(z)$ of positron-emitting isotopes 
of species $i$ along the beam axis $z$ produced per beam particle were calculated with the MCHIT model.
These distributions refer to the secondary nuclei at their stopping points in the medium. 
Then, the depth distribution of the $i$-th
isotope $dN_i(z,t)/dz \equiv F_i(z,t)$ during the irradiation is expressed as a function of time:
\begin{eqnarray}
   {\partial F_i(z,t) \over \partial t} = J f_i(z) - \lambda_i F_i(z,t) 
                                          & {\rm~~~for~~~} & t_j - \tau_s \leq t < t_j~,      \label{EqSpill}\\    
   {\partial F_i(z,t) \over \partial t} = - \lambda_i F_i(z,t)
                                          & {\rm~~~for~~~} & t_j \leq t < t_j + \tau_p~,      \label{EqPause}
\end{eqnarray}
where $\lambda_i$ is the decay constant of the $i$-th isotope, $\lambda_i=\ln(2)/T_{1/2}^i$,
where $T_{1/2}^i$ is the half-life of the $i$-th isotope,
and $t_j \equiv \tau_s + (\tau_p+\tau_s)(j-1)$ is the time when the $j$-th spill ends, $j=1,...,N$. 
Eqs.(\ref{EqSpill}) and (\ref{EqPause}) describe the production and decay of the 
$i$-th isotope during the $j$-th spill and  $j$-th pause, respectively. 
After the irradiation $i$-th isotope decays exponentially:
\begin{equation}
   F_i(z,t) = F_i(z,t_N)\exp( -\lambda_i (t-t_N) ) {\rm~~~for~~~} t \geq t_N~.         \label{ExpDecay}
\end{equation}
The system (\ref{EqSpill}),(\ref{EqPause}) can be solved recursively 
(c.f. Parodi \etal 2002\hide{~\cite{Parodi:etal:2002}}):
\begin{eqnarray}
   F_i(z,t_j) &=& F_i(z,t_{j-1})\exp(-\lambda_i(\tau_p+\tau_s))                          \nonumber \\ 
              &+& F_i(z,t_1) {\rm~~~for~~~} j=2,...,N~,                            \label{EqReq1} \\
   F_i(z,t_1) &=& { J f_i(z) \over \lambda_i } ( 1 - \exp( -\lambda_i \tau_s ) )~.    \label{EqReq2}
\end{eqnarray}
This gives the following expression for the depth distribution of the $i$-th isotope at the end of
$j$-th spill:
\begin{equation}
   F_i(z,t_j) = F_i(z,t_1) \sum_{n=0}^{j-1} \exp( -\lambda_i (\tau_p+\tau_s) n ) 
                            {\rm~~~for~~~} j=1,...,N~.                             \label{SolReq}
\end{equation}

Since the measurements of the $\beta^+$-activity during the irradiation were performed only in pauses
between subsequent spills, the total number of $\beta^+$-decays per unit depth
during the irradiation is
\begin{equation}
   {dN_{\beta^+} \over dz} = \sum_i ( 1 - \exp( -\lambda_i \tau_p ) ) \sum_{j=1}^{N-1} F_i(z,t_j)~.
                                                                                    \label{Acc_d_i}
\end{equation}
After the irradiation the measurements were performed in the time interval from $t_{st}$ to $t_{fin}$
continuously, thus
\begin{equation}
{dN_{\beta^+} \over dz} = \sum_i F_i(z,t_N) [ \exp(-\lambda_i(t_{st}-t_N))-\exp(-\lambda_i(t_{fin}-t_N)) ]~.  \label{Acc_a_i}
\end{equation}

The half-life times of the isotopes included in our analysis and listed in the next section are much longer 
than the spill duration and the pause between spills: $(\tau_s+\tau_p)\lambda_i \ll 1$.
Under this condition Eqs.~(\ref{Acc_d_i}) and (\ref{Acc_a_i}) can be simplified.
The number of $\beta^+$-decays per unit length during the irradiation becomes
\begin{equation}
   {dN_{\beta^+} \over dz} \simeq \sum_i \bar J f_i(z) { \tau_p \over \tau_p+\tau_s }
                                         [ t_N - {1 \over \lambda_i} 
                                                 ( 1 - \exp( -\lambda_i t_N ) ) ]~,  \label{Acc_d_i_simpl}
\end{equation}
while after the irradiation
\begin{eqnarray}
   {dN_{\beta^+} \over dz} &\simeq& \sum_i {\bar J f_i(z) \over \lambda_i} ( 1 - \exp( -\lambda_i t_N ) )
                                                                             \nonumber \\
                           &\times& [ \exp(-\lambda_i(t_{st}-t_N))
                                     -\exp(-\lambda_i(t_{fin}-t_N))]~.        \label{Acc_a_i_simpl}  
\end{eqnarray}
Here $\bar J \equiv J \tau_s / (\tau_p+\tau_s)$ is the average beam intensity calculated for the whole
irradiation period.

\section{Comparison of numerical results with experimental data}\label{comparison}

\subsection{Depth distributions of $\beta^+$-activity}\label{depth-distros}

The spatial distributions of the $\beta^+$-activity induced by 207.92 A MeV $^3$He
and 337.5 A MeV $^{12}$C in various phantoms were measured by 
Fiedler \etal 2006\hide{~\cite{Fiedler:etal:2006}}. In order to validate the
MCHIT model with these data we performed calculations for   
graphite ($9\times9\times15$ cm$^3$, $\rho=1.795$ g cm$^{-3}$), water with an admixture of gelatine   
(H$_{66.2}$O$_{33.1}$C$_{0.7}$, $9\times9\times30$ cm$^3$, $\rho=1.0$ g cm$^{-3}$)   
and polymethyl methacrylate (PMMA, C$_5$H$_8$O$_2$, $9\times9\times20$ cm$^3$, 
$\rho=1.18$ g cm$^{-3}$) phantoms.

Monte Carlo calculations with the MCHIT model have provided the depth distributions in the phantoms
for the following positron-emitting nuclei with $T_{1/2}>10$~s:
$^{10}$C ($T_{1/2}=19.255$ s), $^{11}$C ($T_{1/2}=20.39$ min), $^{13}$N ($T_{1/2}=9.965$ min),
$^{14}$O ($T_{1/2}=1.177$ min), $^{15}$O ($T_{1/2}=2.04$ min), $^{17}$F ($T_{1/2}=1.075$ min),
$^{18}$F ($T_{1/2}=109.77$ min). It was found that these are the most abundant
positron-emitting nuclei produced by $^{3}$He and $^{12}$C in graphite, water and PMMA.
Much lower yields were found for $^{8}$B, $^{9}$C, $^{12}$N and $^{13}$O. Moreover,
as the latter isotopes have rather short half-life time $T_{1/2}<1$~s, 
they decay during the beam spills and do not produce any significant contribution 
during pauses when Fiedler \etal 2006 performed their measurements.

\begin{figure}[htb]  
\begin{centering}
\includegraphics[width=0.65\columnwidth]{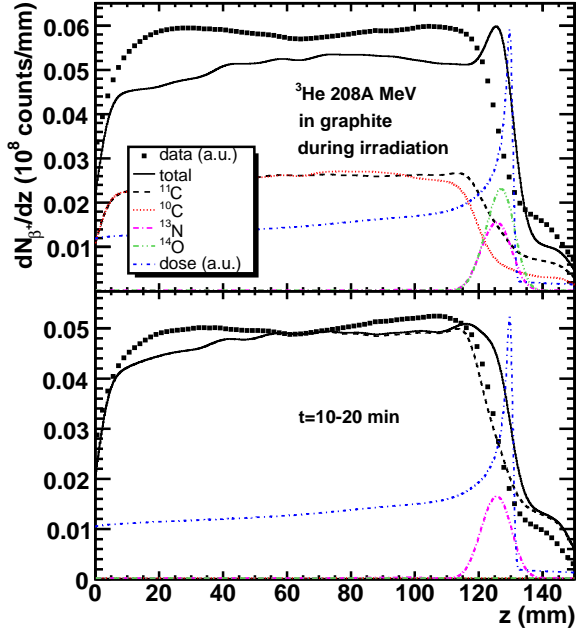}
\end{centering}
\caption{\label{fig:positron_He3_graphite} Depth distribution of
$\beta^+$-activity induced by 207.92 A MeV $^3$He beam in graphite.
The distributions of $\beta^+$-decays counted during irradiation and from 10 to 20 min after it
are shown by solid lines in top and bottom panels, respectively.
Data by Fiedler \etal 2006\hide{~\cite{Fiedler:etal:2006}} are shown by points.
Contributions of specific isotopes and depth-dose distribution are also shown, 
as explained on the legend. 
}
\end{figure}

\begin{figure}[htb]  
\begin{centering}
\includegraphics[width=0.7\columnwidth]{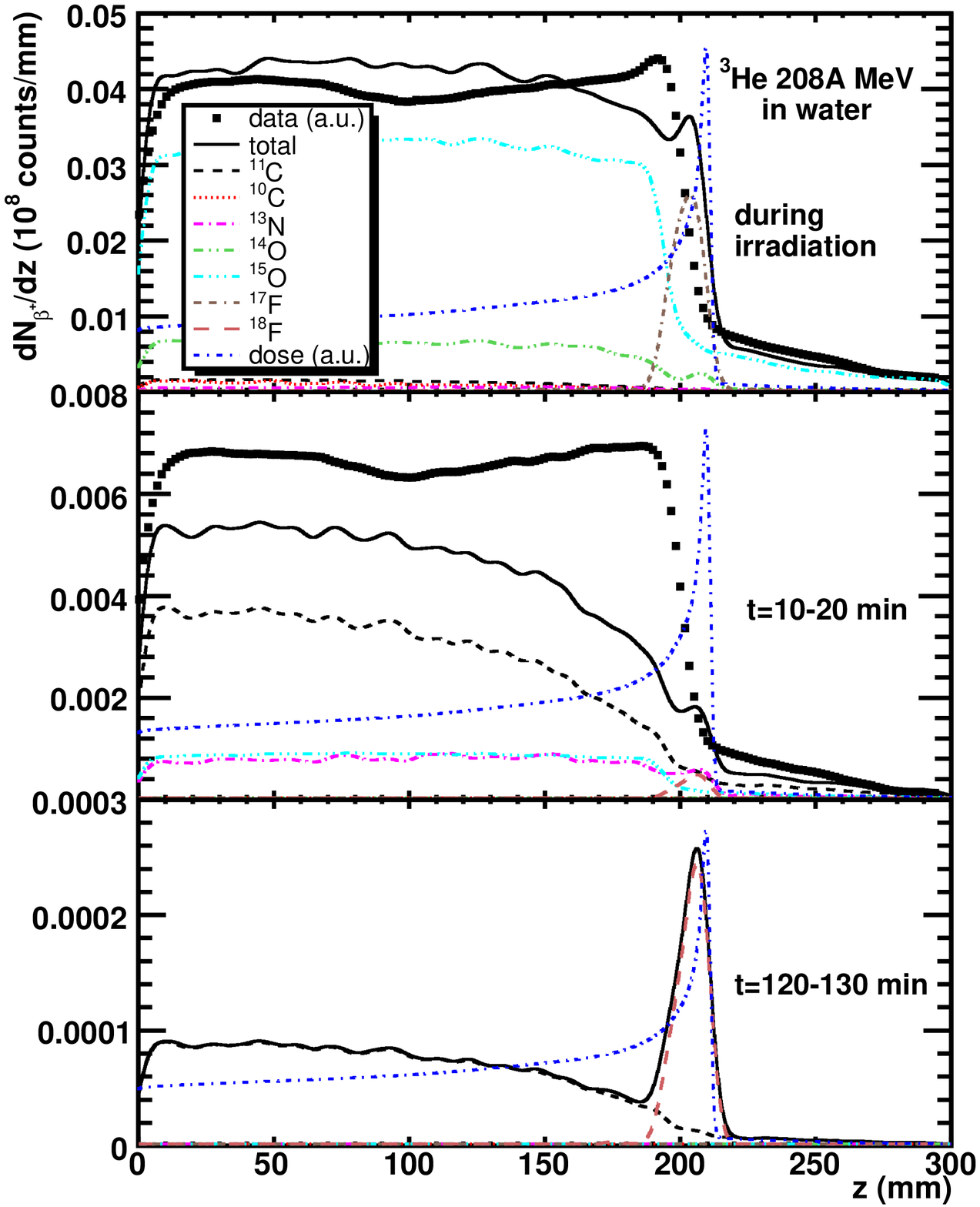}
\end{centering}
\caption{\label{fig:positron_He3_water} Depth distribution of
$\beta^+$-activity induced by 207.92 A MeV $^3$He beam in water.
The distributions of $\beta^+$-decays counted during irradiation, from 10 to 20 min and 
from 120 to 130 min after it are shown by solid lines in top, middle and bottom panels, respectively.
Data by Fiedler \etal 2006\hide{~\cite{Fiedler:etal:2006}} are shown by points.
Contributions of specific isotopes and depth-dose distribution are also shown, 
as explained on the legend. 
}  
\end{figure}

\begin{figure}[htb]  
\begin{centering}
\includegraphics[width=0.7\columnwidth]{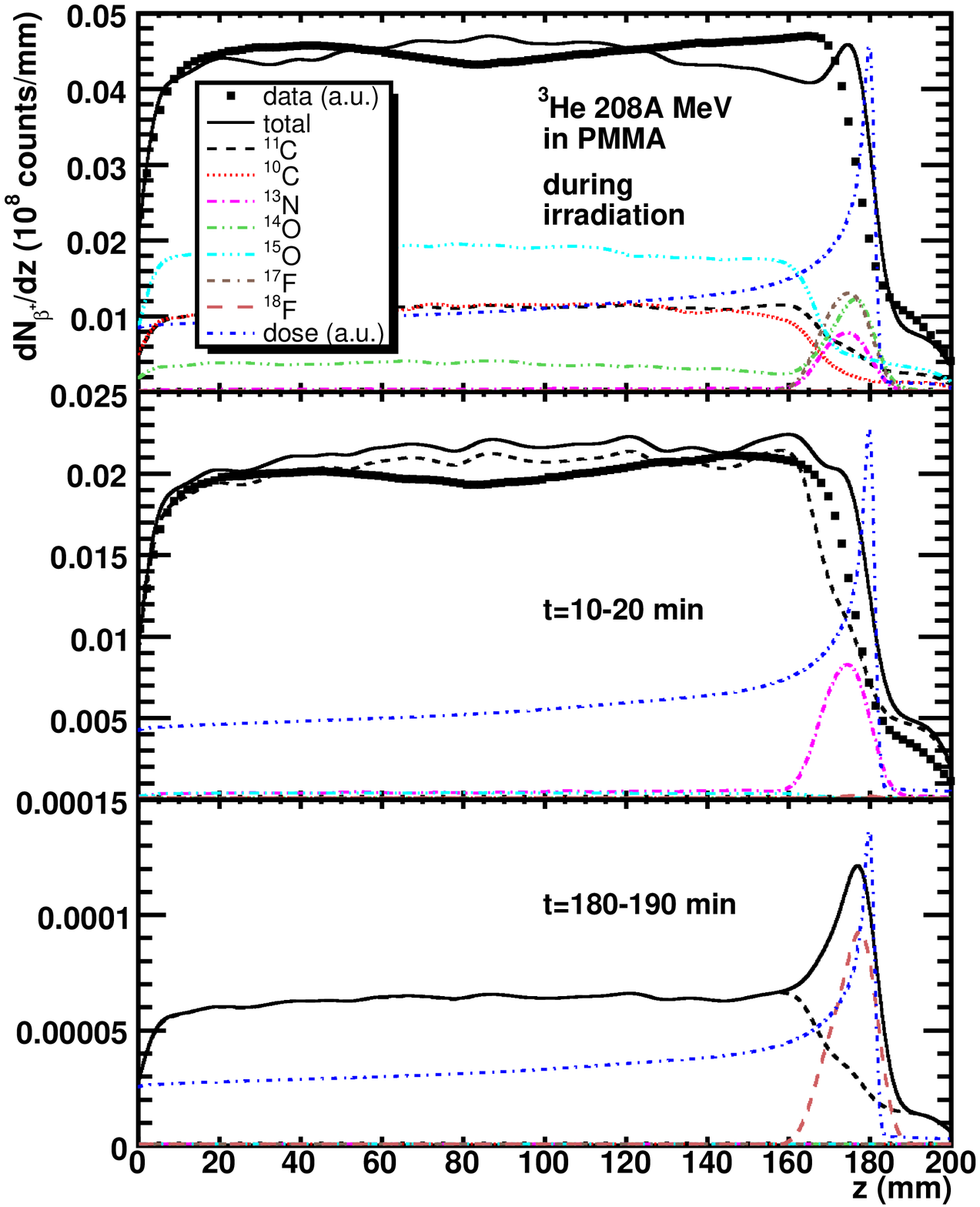}
\end{centering}
\caption{\label{fig:positron_He3_pmma} Same as Fig.~\ref{fig:positron_He3_water},
but for 207.92 A MeV $^3$He beam in PMMA and for  $\beta^+$-decays counted
from 180 to 190 min after irradiation (bottom panel).
}  
\end{figure}

The calculated depth distribution of $\beta^+$-activity for 207.92 A MeV $^3$He beam in
graphite, water and PMMA are shown in Figs.~\ref{fig:positron_He3_graphite}, 
\ref{fig:positron_He3_water} and \ref{fig:positron_He3_pmma} and compared to the data obtained
by Fiedler \etal 2006\hide{~\cite{Fiedler:etal:2006}} 
during irradiation and from 10 to 20 min after it. 
Since measured distributions were presented in arbitrary units,
in Figs.~\ref{fig:positron_He3_graphite}, ~\ref{fig:positron_He3_water} and
~\ref{fig:positron_He3_pmma} they were normalised to the corresponding maxima of 
the calculated $dN_{\beta^+}/dz$ distributions during irradiation. The same weight factor was
applied for plotting the experimental data after irradiation.
  
The $dN_{\beta^+}/dz$ distributions during and after irradiation were calculated according to  
Eqs.(\ref{Acc_d_i}) and (\ref{Acc_a_i}), respectively,  The beam parameters quoted in 
Table~\ref{tab:irradiation_parameters} were used in calculations.  
The calculated $dN_{\beta^+}/dz$ distributions were folded with the Gaussian weigth of
FWHM=8 mm. This width represents the sum of (1) the average distance between a nucleus which emits
a positron and annihilation point $\sim 2$ mm, 
(c.f. Levin and Hoffman 1999)\hide{~\cite{Levin_and_Hoffman:1999}}, and (2)
a finite spatial resolution ($6.5\pm 2$~mm) of the PET scanner used 
by Fiedler \etal 2006\hide{~\cite{Fiedler:etal:2006}}. The shapes of 
the $\beta^+$-activity distributions shown in Figs.~\ref{fig:positron_He3_graphite}
\ref{fig:positron_He3_water} and \ref{fig:positron_He3_pmma}
should be compared with the corresponding depth-dose distributions (given in arbitrary units) 
in order to investigate the correlation between them.  
 
Figures~\ref{fig:positron_He3_graphite}, \ref{fig:positron_He3_water} and \ref{fig:positron_He3_pmma} 
demonstrate also the dependence of the $\beta^+$-activity distributions on the elemental composition 
of the specified target materials. In graphite, which contains only carbon nuclei, 
mostly $^{11}$C and $^{10}$C isotopes are produced by $^3$He via the removal of one or two
neutrons from target $^{12}$C nuclei. As shown in Fig.~\ref{fig:positron_He3_graphite}, these 
nuclei are evenly distributed  within the range of $^3$He projectiles and $^{11}$C is the most 
abundant $\beta^+$-emitter. The MCHIT model predicts a bump near
the Bragg peak due to $^{14}$O and $^{13}$N, which is, however, not visible in the data. 
A smaller bump due to $^{13}$N is also present in the total $\beta^+$-activity distribution 
calculated 10-20 min after irradiation.

A larger set of isotopes is produced by $^3$He in water, see Fig.~\ref{fig:positron_He3_water}.
The most abundant positron-emitting nuclei are $^{15}$O and $^{11}$C.  While
$^{15}$O is produced by the removal of a single neutron from a target $^{16}$O nucleus,
the production mechanism  of $^{11}$C, via the $^{16}$O($^3$He,4p4n)$^{11}$C reaction, is more complicated.
This is reflected in the fact that the overall shape of the total $\beta^+$-activity 
distribution during irradiation is satisfactory reproduced by the MCHIT model since
it is mostly due to $^{15}$O. However, there is a big discrepancy between theory and experiment 
for the time interval 10-20 min after irradiation, as shown in the middle panel of 
Fig.~\ref{fig:positron_He3_water}. It can be explained by the 
deficiency of the model in calculating  $^{16}$O($^3$He,4p4n)$^{11}$C reaction 
which can proceed through various intermediate states and reaction channels.

The MCHIT model predicts a noticeable contribution to the $\beta^+$-activity 
from $^{14}$O during irradiation of water by
$^{3}$He. As the half-life time of $^{14}$O is much shorter than that of $^{11}$C,  
$^{14}$O gives larger contribution during irradiation, 
while $^{11}$C contribution dominates 10-20 min after irradiation. The model also predicts 
a small bump close to the Bragg peak due to $^{17}$F. A similar but somewhat
shifted bump is also seen in the data for 
the total $\beta^+$-activity distribution. The distal slope of the activity distribution can be used for 
determination of the $^{3}$He range in tissues, similar to the proposal by  
Parodi \etal 2002~\hide{~\cite{Parodi:etal:2002}} for therapeutic proton beams.

We have found that a better way to monitor the $^3$He range in tissues can be provided by 
measuring the $\beta^+$-activity at later times, e.g. in the time window of 120-130 min after 
irradiation. The activity distributions for this time interval are also shown  
in Fig.~\ref{fig:positron_He3_water}. Here, the long-living 
$^{18}$F isotope from the $^{16}$O($^3$He,p)$^{18}$F reaction dominates, and the peak in
the activity distribution clearly marks the position of the Bragg peak.

As shown in Fig.~\ref{fig:positron_He3_pmma}, in PMMA
two dominating $\beta^+$-emitters, $^{11}$C and $^{15}$O, are produced by  $^3$He. 
These isotopes are produced in the 
$^{12}$C($^3$He,$\alpha$)$^{11}$C and $^{16}$O($^3$He,$\alpha$)$^{15}$O reactions 
on carbon and oxygen nuclei from PMMA. Due to PMMA chemical composition,  
$^{11}$C is more abundant than $^{15}$O in this material.
The activity distribution in PMMA calculated for later times, e.g. 180-190 min
after irradiation, also has a bump close to the Bragg peak. This activity peak is due to 
$^{18}$F produced in the $^{16}$O($^3$He,p)$^{18}$F reaction.

\begin{figure}[htb]  
\begin{centering}
\includegraphics[width=0.65\columnwidth]{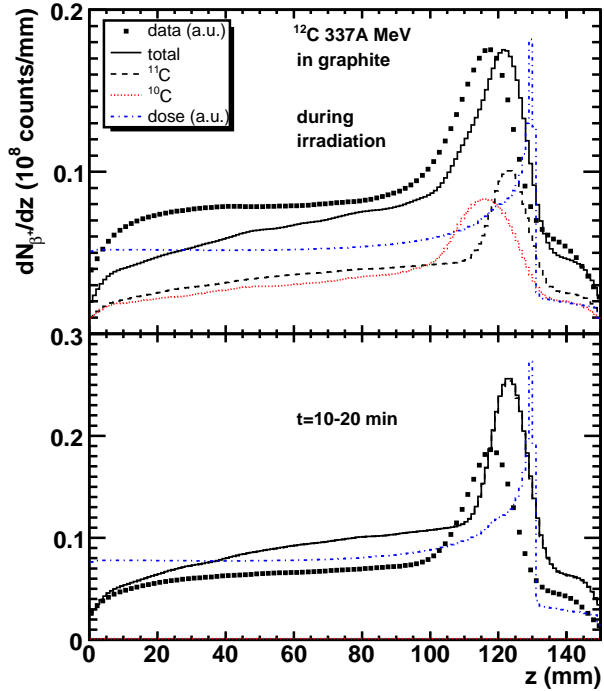}
\end{centering}
\caption{\label{fig:positron_C12_graphite}
Depth distribution of $\beta^+$-activity induced by 337.5 A MeV $^{12}$C beam in graphite.
The distributions of $\beta^+$-decays counted during irradiation and from 10 to 20 min after it
are shown by solid lines in top and bottom panels, respectively.
Data by Fiedler \etal 2006\hide{~\cite{Fiedler:etal:2006}} are shown by points.
Contributions of specific isotopes and depth-dose distribution are also shown, 
as explained on the legend. 
}  
\end{figure}

\begin{figure}[htb]  
\begin{centering}
\includegraphics[width=0.65\columnwidth]{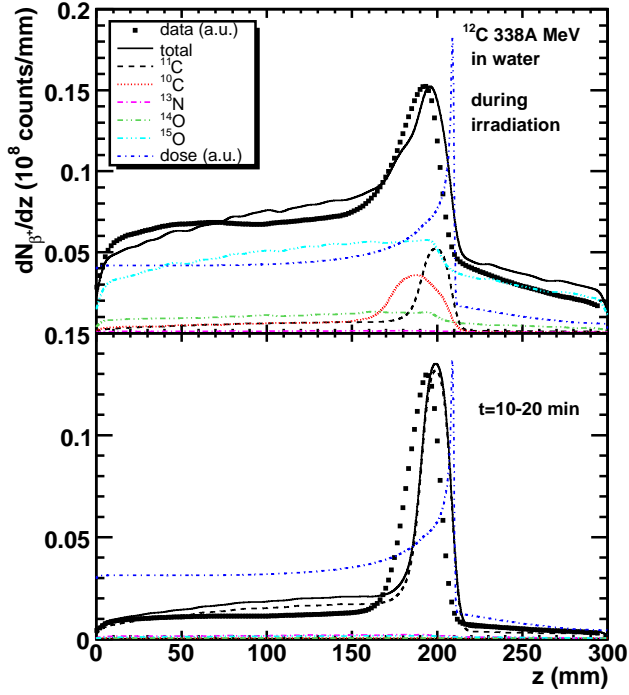}
\end{centering}
\caption{\label{fig:positron_C12_water} Same as Fig.~\ref{fig:positron_C12_graphite},
but for by 337.5 A MeV $^{12}$C beam in water.
}  
\end{figure}

\begin{figure}[htb]  
\begin{centering}
\includegraphics[width=0.65\columnwidth]{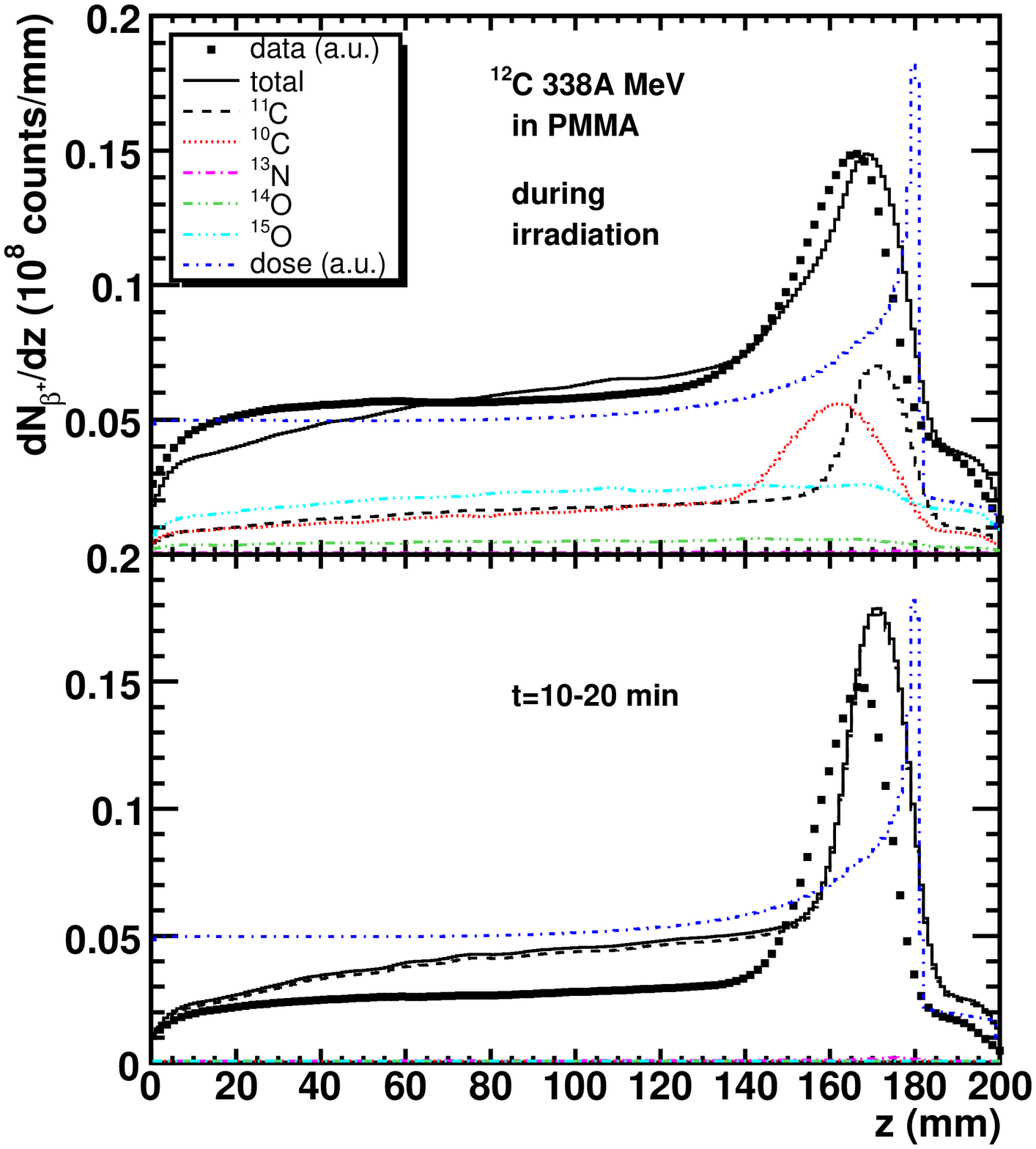}
\end{centering}
\caption{\label{fig:positron_C12_pmma}  Same as  Fig.~\ref{fig:positron_C12_graphite},
but for by 337.5 A MeV $^{12}$C beam in PMMA.
}  
\end{figure}

The MCHIT model is also verified with the $\beta^+$-activity distributions measured by 
Fiedler \etal 2006\hide{~\cite{Fiedler:etal:2006}} for  337.5 A MeV $^{12}$C beam
in graphite, water and PMMA, as shown in Figs.~\ref{fig:positron_C12_graphite}, 
\ref{fig:positron_C12_water} and \ref{fig:positron_C12_pmma}.
In these phantom materials $^{10}$C and $^{11}$C can be produced by
single or double neutron removal from both $^{12}$C projectiles and $^{12}$C target nuclei.     
As a result, the $\beta^+$-activity distribution is characterised by sharp peaks due to 
projectile fragmentation and plateau due to target fragmentation, 
as seen, in particular, in Fig.~\ref{fig:positron_C12_graphite}. 

The overall shapes of the total $\beta^+$-activity distributions 
are satisfactorily described by the MCHIT model. However, as a rule,
the peaks in theoretical distributions are located 5-10~mm deeper 
compared to the experimental ones. This shift is caused by the overestimation of
the $^{11}$C production in the binary cascade model at low energies, as explained in
Section~\ref{discussion}.

As seen in Figs.~\ref{fig:positron_C12_graphite}, \ref{fig:positron_C12_water} 
and \ref{fig:positron_C12_pmma} for all three materials, only $^{11}$C survives within 
10 min after irradiation.
However, during irradiation the contributions from $^{14}$O and $^{15}$O 
are important for water and PMMA. These positron-emitting oxygen isotopes
are also produced beyond the Bragg peak by energetic secondary protons emitted in nuclear 
fragmentation of $^{12}$C beam.

\subsection{Total yields of positron-emitting nuclei produced by $^3$He and $^{12}$C beams}

The MCHIT model can further be validated by confronting calculated total  
production yields of specific positron-emitting nuclei with the yields measured by
Fiedler \etal 2006\hide{~\cite{Fiedler:etal:2006}}. 
In Tables~\ref{tab:isotope_yield_graphite}, \ref{tab:isotope_yield_water} and 
\ref{tab:isotope_yield_pmma} we present the total yields of the most 
abundant isotopes, $^{10}$C, $^{11}$C, $^{13}$N and $^{15}$O, produced by $^3$He and $^{12}$C beams
in graphite, water and PMMA phantoms. The values are given in \% per beam particle.

From inspecting Table~\ref{tab:isotope_yield_graphite} one can conclude that the calculated 
yields of $^{11}$C are in better agreement with experiment than $^{10}$C yields.
The yields of $^{11}$C produced by $^3$He are well described, while the model overestimates 
the production of  $^{10}$C by both beams.    

\begin{table}[htb]
    \caption{\label{tab:isotope_yield_graphite} Computed number of $\beta^+$-emitters 
             (in \% per beam particle) from interactions of $^3$He and $^{12}$C in graphite.
             Data are from Fiedler \etal 2006\hide{~\cite{Fiedler:etal:2006}}.
            }
\begin{indented}
\item[]\begin{tabular}{@{}llllllllll}
\br
        & \multicolumn{2}{c}{$^3$He, 130.03A MeV} & \multicolumn{2}{c}{$^3$He, 166.05A MeV}%
        & \multicolumn{2}{c}{$^3$He, 207.92A MeV} & \multicolumn{2}{c}{$^{12}$C, 337.5A MeV}\\
\mr
        & MCHIT          & Experiment     & MCHIT           & Experiment          & MCHIT   & Experiment%
        & MCHIT          & Experiment \\
\mr 
$^{10}$C& 0.712          & 0.44$\pm$0.07  & 1.07            & 0.69$\pm$0.10       & 1.51    & 0.98$\pm$0.15%
        & 4.56           & 1.88$\pm$0.28  \\
$^{11}$C& 5.28           & 6.9$\pm$1.0    & 7.99            & 10.0$\pm$1.5        & 11.09   & 13.9$\pm$2.1%
        & 34.80          & 24.9$\pm$3.7  \\
$^{13}$N& 0.508          & ---            & 0.460           & ---                 & 0.358   & ---%
        & 0.155          & --- \\
$^{15}$O& 0.0021         & ---            & 0.0022          & ---                 & 0.002   & ---%
        & 0.031          & --- \\
\br
\end{tabular}
\end{indented}
\end{table}

One can also compare the model with the data on $^{13}$N and $^{15}$O production in irradiation of water 
phantoms by $^{3}$He and $^{12}$C. Calculated and measured yields for this case are 
presented in Table~\ref{tab:isotope_yield_water}. The yields of $^{15}$O produced 
by $^3$He are underpredicted by the model by $\sim 30$\% for all $^3$He energies, while 
the production of $^{15}$O and $^{11}$C by $^{12}$C beam is well described. 
On the other hand, the model completely fails in describing $^{13}$N and $^{11}$C production by
$^3$He, as these yields are underestimated by a factor of three. 
We attribute this problem to the complexity of 
$^{16}$O($^3$He,3p3n)$^{13}$N and $^{16}$O($^3$He,4p4n)$^{11}$C reactions.
The ability of the model to predict $^{10}$C yields for $^3$He beams depends on the beam energy.
The calculations agree better with the data at 207.92 A MeV than at 130.03 and 166.05 A MeV.  

\begin{table}[htb]
    \caption{\label{tab:isotope_yield_water} Computed number of $\beta^+$-emitters 
             (in \% per beam particle) from interactions of $^3$He and $^{12}$C in water.
             Data are from Fiedler \etal 2006\hide{~\cite{Fiedler:etal:2006}}.
            }
\begin{indented}
\item[]\begin{tabular}{@{}llllllllll}
\br
        & \multicolumn{2}{c}{$^3$He, 130.03A MeV} & \multicolumn{2}{c}{$^3$He, 166.05A MeV}%
        & \multicolumn{2}{c}{$^3$He, 207.92A MeV} & \multicolumn{2}{c}{$^{12}$C, 337.5A MeV} \\
\mr
        & MCHIT          & Experiment     & MCHIT           & Experiment          & MCHIT   & Experiment%
        & MCHIT          & Experiment \\
\mr 
$^{10}$C& 0.039          & 0.21$\pm$0.03  & 0.071           & 0.19$\pm$0.03       & 0.118   & 0.18$\pm$0.04%
        & 1.93           & 0.78$\pm$0.12  \\
$^{11}$C& 0.415          & 1.85$\pm$0.28  & 0.784           & 2.49$\pm$0.37       & 1.28    & 3.23$\pm$0.48%
        & 13.1           & 12.6$\pm$1.9   \\
$^{13}$N& 0.154          & 0.49$\pm$0.07  & 0.235           & 0.80$\pm$0.12       & 0.321   & 1.02$\pm$0.15%
        & 1.02           & 2.40$\pm$0.36  \\
$^{15}$O& 2.30           & 4.40$\pm$0.66  & 3.75            & 6.29$\pm$0.94       & 5.65    & 8.29$\pm$1.24%
        & 15.3           & 14.6$\pm$2.2 \\
\br
\end{tabular}
\end{indented}
\end{table}

Calculations and experimental data for PMMA phantoms irradiated by
$^3$He and $^{12}$C are presented in Table~\ref{tab:isotope_yield_pmma}.
The calculated  yields of $^{10}$C and $^{13}$N produced by $^3$He in this material are
well described by the MCHIT model, while the yields of most abundant  $^{11}$C and $^{15}$O  
are underestimated by $\sim 30$\%. The model is quite successful in describing
$^{13}$N, $^{11}$C and $^{15}$O produced by  $^{12}$C beam, but 
the production of $^{10}$C is overestimated.

\begin{table}[htb]
    \caption{\label{tab:isotope_yield_pmma} Computed number of $\beta^+$-emitters 
             (in \% per beam particle) from interactions of $^3$He and $^{12}$C in PMMA.
             Data are from Fiedler \etal 2006\hide{~\cite{Fiedler:etal:2006}}.
            }
\begin{indented}
\item[]\begin{tabular}{@{}llllllllll}
\br
        & \multicolumn{2}{c}{$^3$He, 130.03A MeV} & \multicolumn{2}{c}{$^3$He, 166.05A MeV}%
        & \multicolumn{2}{c}{$^3$He, 207.92A MeV} & \multicolumn{2}{c}{$^{12}$C, 337.5A MeV} \\
\mr
        & MCHIT          & Experiment     & MCHIT           & Experiment          & MCHIT   & Experiment%
        & MCHIT          & Experiment \\
\mr 
$^{10}$C& 0.398          & 0.38$\pm$0.06  & 0.607           & 0.53$\pm$0.08       & 0.862   & 0.68$\pm$0.10%
        & 3.30           & 1.64$\pm$0.25  \\
$^{11}$C& 2.89           & 4.70$\pm$0.71  & 4.47            & 7.05$\pm$1.06       & 6.26    & 9.64$\pm$1.45%
        & 23.5           & 22.0$\pm$3.3   \\
$^{13}$N& 0.332          & 0.23$\pm$0.05  & 0.337           & 0.28$\pm$0.06       & 0.321   & 0.44$\pm$0.09%
        & 0.425          & 0.63$\pm$0.13   \\
$^{15}$O& 0.87           & 1.53$\pm$0.23  & 1.39            & 2.35$\pm$0.35       & 2.06    & 3.19$\pm$0.48%
        & 5.10           & 5.14$\pm$0.77  \\
\br
\end{tabular}
\end{indented}
\end{table}

In summary, the yields of the most abundant isotopes $^{11}$C and $^{15}$O 
produced by $^{12}$C in water and PMMA are very well described by the MCHIT model, 
see Tables~\ref{tab:isotope_yield_water} and \ref{tab:isotope_yield_pmma}.
However, the yields of $^{11}$C and $^{15}$O, which are 
abundantly produced also by $^3$He in water and PMMA, are underpredicted at all  $^3$He
energies. This means that there is a room for improvement of the GEANT4 nuclear reaction models with
respect to $^3$He-induced reactions.

\section{Calculations of $\beta^+$-activity distributions in tissues}\label{tissues}

\begin{figure}[htb]  
\begin{centering}
\includegraphics[width=0.7\columnwidth]{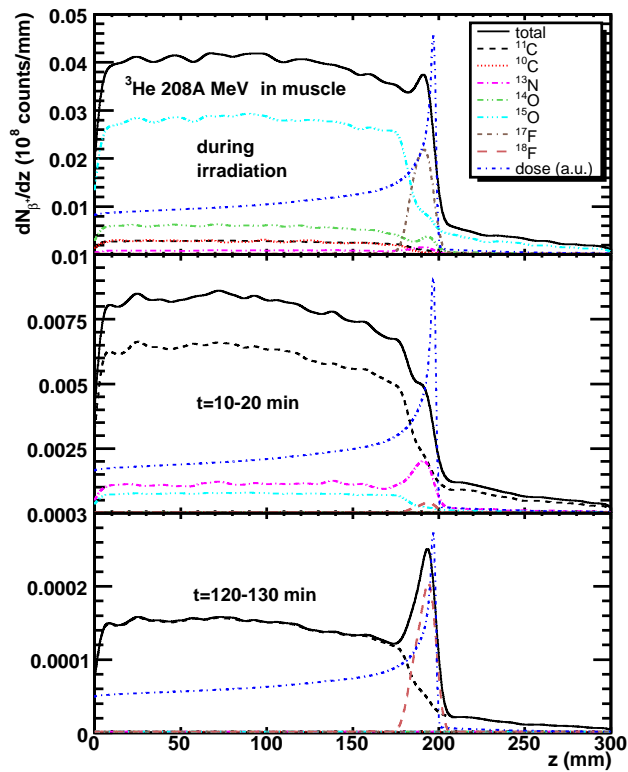}
\end{centering}
\caption{\label{fig:positron_He3_muscle}
Depth distribution of
$\beta^+$-activity induced by 207.92 A MeV $^3$He beam in muscle tissue.
The distributions of $\beta^+$-decays counted during irradiation, from 10 to 20 min and 
from 120 to 130 min after it are shown by solid lines in top, middle and bottom panels, 
respectively. Contributions of specific isotopes and depth-dose distribution are also shown, 
as explained on the legend. 
}  
\end{figure}

\begin{figure}[htb]  
\begin{centering}
\includegraphics[width=0.7\columnwidth]{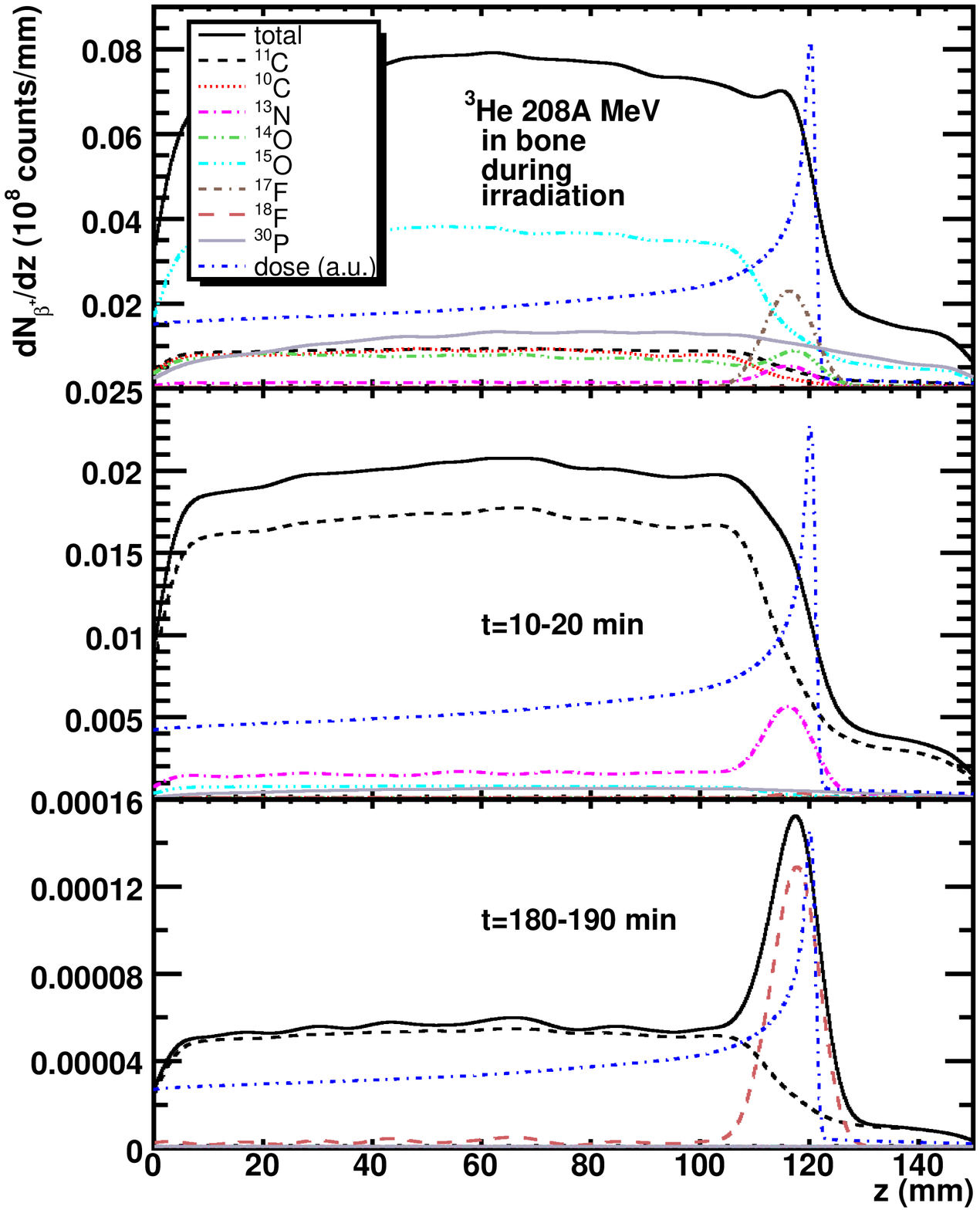}
\end{centering}
\caption{\label{fig:positron_He3_bone} Same as Fig.~\ref{fig:positron_He3_muscle},
but for 207.92 A MeV $^3$He beam in compact bone tissue and for  $\beta^+$-decays counted
from 180 to 190 min after irradiation (bottom panel).
}  
\end{figure}

As demonstrated above, the total yields and spatial distributions of $\beta^+$-activity produced
by therapeutic beams depend essentially on the elemental compositions of target materials. 
Therefore, for studying the feasibility of the PET monitoring method in real tissues irradiated
with $^3$He we have performed calculations for two homogeneous phantoms with elemental composition 
similar to muscle 
($9\times9\times30$ cm$^3$, $\rho=1.061$ g cm$^{-3}$) and compact bone 
($9\times9\times15$ cm$^3$, $\rho=1.850$ g cm$^{-3}$). The elemental
composition was taken in the following mass fractions:
H - 10.2 \%, C - 14.3\%, N - 3.4 \%, O - 71\%, Na - 0.1 \%,
P - 0.2\%, S - 0.3 \%, Cl - 0.1 \%, K - 0.4\% for muscle tissue, and 
H - 6.4 \%, C - 27.8\%, N - 2.7 \%, O - 41\%, Mg - 0.2 \%,
P - 7\%, S - 0.2 \%, Ca - 14.7 \% for compact bone tissue.
The beam parameters in calculations of $^3$He irradiation of muscle (bone) were taken the same as for
$^3$He beam in water (graphite), as listed in Table~\ref{tab:irradiation_parameters}. 
The calculated depth distributions of $\beta^+$-activity in tissues irradiated by 207.92 A MeV $^3$He
are shown in Figs.~\ref{fig:positron_He3_muscle} and \ref{fig:positron_He3_bone}.

Due to the presence of Na, P, S, Cl, Ca, K and Mg in muscle and bone tissues, 
$^{19}$Ne ($T_{1/2}=17.22$~s),   
$^{21}$Na ($T_{1/2}=22.49$~s), and $^{30}$P ($T_{1/2}=2.498$ min) can also be produced in fragmentation
reactions in addition to the isotopes analysed in the previous sections. 
However, only negligible yields of $^{19}$Ne and $^{21}$Na are predicted by the MCHIT model 
and they can be safely neglected. The $\beta^+$-activity distributions in muscle are similar to those
in water (c.f. Fig.~\ref{fig:positron_He3_water}). In bone tissue, however, $^{30}$P is
additionally produced by the $^{31}$P target fragmentation. In fact, both $^{30}$P and
$^{15}$O noticeably contribute to the $dN_{\beta^+}/dz$ after the distal edge of the Bragg peak,
as shown in Fig.~\ref{fig:positron_He3_bone}. This is similar to $^{11}$C distribution 
in graphite and PMMA irradiated by $^3$He (see Figs.~\ref{fig:positron_He3_graphite} 
and \ref{fig:positron_He3_pmma} where also a tail of the $\beta^+$-activity is present beyond
the Bragg peak). 

Only $^{11}$C and $^{18}$F survive in muscle and bone tissues at later time after irradiation
The presence of $^{18}$F opens a new way to monitor the $^3$He range in tissues, as
the $^{18}$F peak (see the bottom panels of Figs.~\ref{fig:positron_He3_muscle}  and
\ref{fig:positron_He3_bone}) clearly mark the position of the Bragg peak.

\section{PET monitoring with low energy proton, $^3$He and $^{12}$C beams}\label{LowEnergy}

The nuclear pick-up reactions leading to the production of $^{14}$O, $^{17}$F and
$^{18}$F nuclei have the maximal cross sections at low energies. 
In this energy regime the velocities of projectile nuclei are comparable to the velocities 
of intranuclear nucleons due to Fermi motion. This gives optimum conditions for transferring 
nucleons from one collision partner to another during their collision and enhances the 
production of  $^{14}$O, $^{17}$F and $^{18}$F.   

\begin{figure}[htb]  
\begin{centering}
\includegraphics[width=0.7\columnwidth]{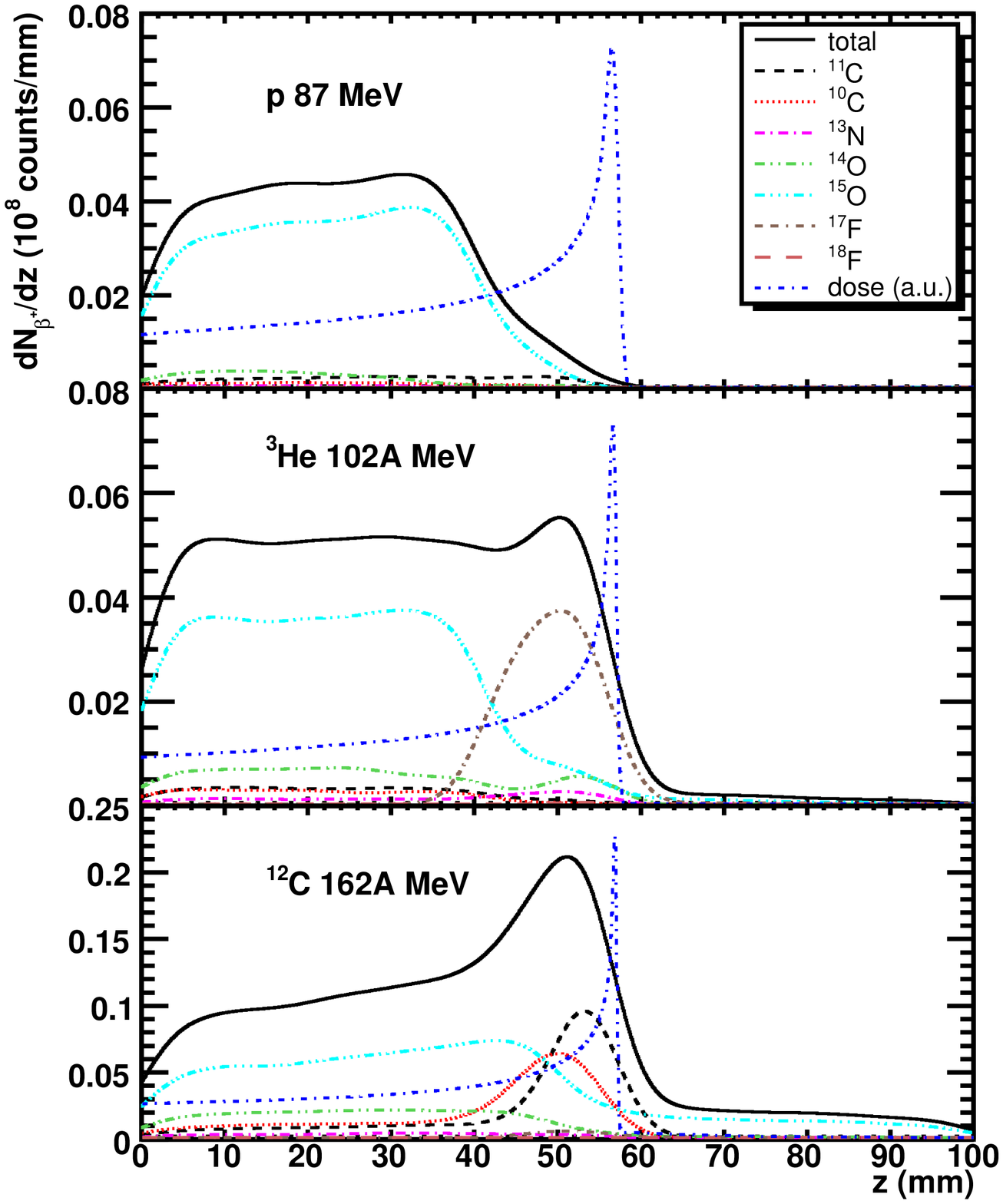}
\end{centering}
\caption{\label{fig:p_He3_C12_Muscle} Depth distribution of
of $\beta^+$-activity during irradiation by 87 MeV protons (top panel), 
102 A MeV $^{3}$He (middle panel) and 162 A MeV $^{12}$C beam in muscle tissue.
}  
\end{figure}

It is instructive to consider the distributions of $\beta^+$-activity in muscle produced by
various low-energy beams during irradiation, as shown in Fig.~\ref{fig:p_He3_C12_Muscle}. 
In these calculations the time structure of all three beams was assumed 
the same as for graphite irradiation by $^3$He, see Table.~\ref{tab:irradiation_parameters}. 
It is expected that nuclear transfer reactions are more important at low energies, 
while nuclear fragmentation reactions contribute less because they have certain 
energy thresholds.

As one can see in  Fig.~\ref{fig:p_He3_C12_Muscle}, the distribution of positron emitting nuclei
produced by low-energy 87 MeV proton beam is almost uniform and poorly correlated with the position of the Bragg peak. 
In fact, the Bragg peak in the depth-dose distribution is located at the region with
a negligible $\beta^+$-activity. A similar distribution is predicted for $^{15}$O produced by
102 A MeV $^3$He. However, $^{17}$F nuclei are additionally produced by $^3$He in nuclear charge pick-up
reactions close to the Bragg peak. This makes the total distribution of $\beta^+$-activity for
$^3$He more suitable for the determination of the $^{3}$He range in tissues by the PET method, as the
distal end of the $\beta^+$-activity distribution marks clearly the position of the Bragg peak. 
It is advisable to perform the PET measurements with low-energy  $^3$He beams to
quantify the contribution of charge pick-up reactions.  
The $\beta^+$-activity distribution produced by 162 A MeV $^{12}$C is also suitable for 
PET monitoring due to the presence of broad peaks associated with $^{10}$C and $^{11}$C nuclei.

\section{Reliability of calculational results}\label{discussion}

We discuss the discrepancies between the MCHIT results and the experimental data
by Fiedler \etal 2006. The calculations agree with the   
data on the total yields of $^{10}$C, $^{11}$C,  $^{13}$N and $^{15}$O
in graphite, water and PMMA at $\sim 30-50$\% accuracy level, see Tables 
\ref{tab:isotope_yield_graphite}-\ref{tab:isotope_yield_pmma}. However,
the agreement with the activity distribution measured 10-20 min after
irradiation of water by  207.92 A MeV  $^3$He is poor, see the middle panel of 
Fig.~\ref{tab:isotope_yield_water}. We identify this discrepancy with the poor description of
$^{16}$O($^3$He,4p4n)$^{11}$C reaction, and we conclude that it has to be improved in GEANT4.
In this reaction a compound nucleus $^{19}$Ne can be created leading to a larger
longitudinal momentum transfer as compared with the direct mechanism. 
Therefore, the $^{11}$C nuclei produced in decays of $^{19}$Ne will have on average a 
larger longitudinal momentum and will stop closer to the Bragg peak. Therefore, the agreement in shapes of
the calculated and measured distributions of $^{11}$C nuclei can be possibly improved
by taking into account the formation of $^{19}$Ne.   

The model was also confronted with the measured activity distributions from 337.5 A MeV $^{12}$C 
in graphite, water and PMMA. 
The largest discrepancy between calculations and data obtained 10-20 min after irradiation 
was found for the graphite phantom, see the bottom panel of Fig.~\ref{fig:positron_C12_graphite}.
To identify the origin of this discrepancy the cross section $\sigma(^{11}$C) of
the $^{12}$C($^{12}$C,n)$^{11}$C reaction was calculated with the GEANT4 toolkit. 
This was done following the expression:
\begin{equation}
   \sigma(^{11}{\rm C}) = {d N_{^{11}{\rm C}} \over d z}_{|z=0} n^{-1}~,            \label{Sigma_C11}
\end{equation}
where ${d N_{^{11}{\rm C}} / d z}_{|z=0}$ is the number of $^{11}$C nuclei per incident $^{12}$C ion
per unit length at the entrance point $z=0$ to the graphite. In fact, we have averaged
$d N_{^{11}{\rm C}} / d z$ over the region $0 < z < 1$ mm. The $^{12}$C concentration in graphite is 
$n = 9. \cdot 10^{22}$ cm$^{-3}$. 
The obtained results are compared with the data by Yashima \etal 2003, 
2004~\hide{~\cite{Yashima:etal:2003,Yashima:etal:2004}} in Table \ref{tab:Sigma_C11}.
One can see that the MCHIT model overpredicts the cross section $\sigma(^{11}{\rm C})$ by 30-60\%. The ratio
$\sigma_{^{11}{\rm C}}^{\rm GEANT4}/\sigma_{^{11}{\rm C}}^{\rm exp}$ seems to grow at lower  
beam energies. This 
explains the shift of the peak of the calculated activity distribution to larger $z$ with respect to 
the data.

\begin{table}[htb]
    \caption{\label{tab:Sigma_C11} Cross section of the reaction $^{12}$C($^{12}$C,X)$^{11}$C
calculated using GEANT4 in comparison to the data from Yashima \etal 2003, 2004
\hide{~\cite{Yashima:etal:2003,Yashima:etal:2004}}. The error bars on theoretical results
are pure statistical.
            }
\begin{indented}
\item[]\begin{tabular}{@{}lllllll}
\br
        & E/A, MeV & $\sigma_{^{11}{\rm C}}^{\rm GEANT4}$, mb & $\sigma_{^{11}{\rm C}}^{\rm exp}$, mb   \\
\mr
        & 100      & 144$\pm$14                    & 88.3$\pm$3.2                \\
        & 230      & 106$\pm$17                    & 79.0$\pm$7.9                \\
        & 400      & 100$\pm$10                    & 68.6$\pm$2.5                \\

\br
\end{tabular}
\end{indented}
\end{table}

The production of $^{18}$F by 36-40 MeV $^3$He ions in water was studied by
Fitschen \etal 1977~\hide{\cite{Fitschen:etal:1977}} and by   
Knust and Machulla 1983~\hide{\cite{Knust:and:Machulla:1983}}.
They reported 20 mCi/$\mu$A and 19 mCi/$\mu$A activity of $^{18}$F after
3 and 2.5 hours of irradiation with $^3$He beam, respectively.
The activity calculated by the MCHIT model for similar irradiation conditions amounts to 
18.7 mCi/$\mu$A, which is in very good agreement with the experimental yields. This analysis
shows that the low-energy nuclear data play an important role in designing reliable models
for heavy-ion cancer therapy.

\section{Summary and conclusions}\label{summary}

We have considered nuclear reactions induced by $^3$He and $^{12}$C beams 
in tissue-like materials from the view point of their suitability for PET monitoring. 
As found, in addition to nuclear fragmentation reactions of projectile and target nuclei, leading
to creation of $^{10,11}$C  and $^{14,15}$O, the contributions from several nuclear pick-up  
reactions, $^{12}$C($^3$He,X)$^{13}$N, $^{12}$C($^3$He,n)$^{14}$O, 
$^{16}$O($^3$He,X)$^{17}$F and $^{16}$O($^3$He,p)$^{18}$F must be taken into account.
The pick-up of nucleons is quite efficient at low collision energies when the relative velocities of nuclei
are comparable to the characteristic velocities of intranuclear nucleons due to their Fermi motion. 
On the other hand, it is known that this is the region of ion-beam energies with the maximum of relative
biological effectiveness (RBE).

As pointed out a long time ago by Osgood \etal 1964, Cirilov \etal 1966 and Hahn and Ricci 1966,   
\hide{~\cite{Osgood:etal:1964,Cirilov:etal:1966,Hahn:and:Ricci:1966}}
there exist different mechanisms of pick-up  reactions. Indeed,
the proton stripping reaction $^{12}$C($^3$He,d)$^{13}$N is a direct reaction, 
while the $^{12}$C($^3$He,n)$^{14}$O process goes through the formation of $^{15}$O compound nucleus. 
In both cases, the reaction cross section increases at low projectile
energies. As follows from our calculations, $^{13}$N and $^{14}$O are concentrated near the Bragg peak of 
$^3$He in graphite and PMMA (Figs.~\ref{fig:positron_He3_graphite} and \ref{fig:positron_He3_pmma}),
while the distributions of $^{10}$C, $^{11}$C and $^{15}$O are flat
because such nuclides are resulting from target fragmentation reactions. 

The process $^{16}$O($^3$He,p)$^{18}$F is of particular interest since $^{18}$F is a long living isotope 
($T_{1/2}=109.77$ min), making it suitable for off-line monitoring of $\beta^+$-activity.
 According to Hahn and Ricci 1966,\hide{~\cite{Hahn:and:Ricci:1966}} 
the cross section of this reaction has a maximum $436\pm44$ mb at $E_{lab}=6.3$ MeV and drops at higher beam energies. 
As demonstrated above, 2-3 hours after irradiation by $^3$He a peak in the $\beta^+$-activity 
distribution due to $^{18}$F is developed in water and PMMA. The position of this peak well matches the position 
of the Bragg peak. Counting statistics in the experiment by 
Fiedler \etal 2006\hide{~\cite{Fiedler:etal:2006}} should, in principle, allow to identify this peak.  
Since the measured biological wash-out time is about of 91-124~min in muscle 
(Tomitani \etal 2003)\hide{~\cite{Tomitani:etal:2003}}, i.e. comparable to physical 
half-life of $^{18}$F, the biological wash-out should not 
drastically decrease PET signal even for water-dominated tissues.
In bone tissue PET signal is expected to be robust and survive 2-3 hours after irradiation. 
Alternatively, the biological wash-out of 
$\beta^+$-activity at later times ($\ge 1$ hour) may be studied with $^{18}$F.

\ack
This work was supported by Siemens Medical Solutions.
We are grateful to Prof. Hermann Requardt for stimulating discussions. 
The discussions with  Dr. Thomas Haberer and Dr. Dieter Schardt 
are gratefully acknowledged. 
We are indebted to Prof. Wolfgang Enghardt and Dr. Fine Fiedler for discussions and for 
providing us with the tables of their experimental data, and their compilation of experimental
data on $^3$He-induced nuclear reactions.

% File for Harvard system

\References 

%*\bibitem{Agostinelli:etal:2003} 
\item[] Agostinelli S \etal  (GEANT4 Collaboration) 2003 GEANT4: 
A simulation toolkit {\it Nucl. Instrum. Methods} A {\bf 506} 250-303

%*\bibitem{Allison:etal:2006}
\item[] Allison J \etal (GEANT4 Collaboration) 2006 
GEANT4 developments and applications
{\it IEEE Trans. Nucl. Sci.} {\bf 53} 270-8 

%*\bibitem{Amaldi:and:Kraft:2005} 
\item[] Amaldi U and Kraft G 2005
Radiotherapy with beams of carbon ions
{\it Rep. Prog. Phys.} {\bf 68} 1861-82

%*\bibitem{Amaldi:2004}
\item[]   Amaldi U  2004
 CNAO--The Italian Centre for Light-Ion Therapy
{\it Radiother. Oncol.} {\bf 73}  S191-201

%*\bibitem{Bajard:etal:2004}
\item[]  Bajard M, De Conto J M and Remillieux J 2004
Status of the "ETOILE" project for a French hadrontherapy centre   
{\it Radiother. Oncol.} {\bf 73} S211-5

%*\bibitem{Bennett:etal:1975}
\item[] Bennett G W, Goldberg A C, Levine G S, Guthy J, Balsamo J and Archambeau J O  1975
Beam localization via O-15 activation in proton-radiation therapy
{\it Nucl. Instr. Methods} {\bf 125} 333-8

%*\bibitem{Bennett:etal:1978}
\item[] Bennett G W, Archambeau J O, Archambeau B E, Meltzer J I and Wingate C L  1978 
Visualization and transport of positron emission from proton activation in vivo 
{\it Science} {\bf 200} 1151-3

%*\bibitem{Bondorf:etal:1995} 
\item[] Bondorf J P, Botvina A S, Iljinov A S, Mishustin I N and Sneppen K  1995 
Statistical multifragmentation of nuclei {\it Phys. Rept.} {\bf 257} 133-221

%*\bibitem{Castro:etal:2004}
\item[] Castro J R, Petti P L, Blakely E A and Daftari I K 2004
Particle radiation therapy 
{\it Textbook of Radiation Oncology} (Saunders, Elsevier Inc.)
ed Leibel S A and Phillips T L 
pp 1547-68 

%\bibitem{Cirilov:etal:1966}
\item[]  Cirilov S D, Newton J O and Schapira J P 1966
Total cross sections for the reaction $^{12}$C($^3$He,$\alpha$)$^{11}$C 
and $^{12}$C($^3$He,n)$^{14}$O
{\it Nucl. Phys.} {\bf 77} 472-6.

%*\bibitem{Enghardt:etal:2004}
\item[]  Enghardt W, Crespo P, Fiedler F, Hinz R, Parodi K, Pawelke J, and P\"{o}nisch F 2004
Charged hadron tumour therapy monitoring by means of PET
{\it Nucl. Instrum. Methods} A {\bf 525} 284-8

%*\bibitem{Enghardt:etal:1992}
\item[] Enghardt W, Fromm W D, Geissel H, Keller H, Kraft G, Magel A, Manfrass P, 
Munzenberg G, Nickel F, Pawelke J, Schardt D, Scheidenberger C and Sobiella M  1992
The spatial-distribution of positron-emitting nuclei generated by relativistic light-ion
beams in organic-matter
{\it Phys. Med. Biol.} {\bf 37}  2127-31

%*\bibitem{Fiedler:etal:2006}
\item[] Fiedler F, Crespo P, Parodi K, Sellesk M and Enghardt W 2006
The feasibility of in-beam PET for therapeutic beams of $^3$He
{\it IEEE Trans. Nucl. Sci.} {\bf 53} 2252-9 

%*\bibitem{Fitschen:etal:1977}
\item[] Fitschen J, Beckmann R, Holm U and Neuert H 1977
Yield and production of $^{18}$F by $^3$He irradiation of water
{\it Int. J. Appl. Radiat. Isot.} {\bf 28} 781-884

%*\bibitem{Folger:etal:2004} 
\item[] Folger  G, Ivanchenko V N and Wellisch J P  2004 The Binary Cascade - 
nucleon-nuclear reactions {\it Eur. Phys. J.} A {\bf 21} 407-17

%*\bibitem{Furusawa:etal:2000}
\item[] Furusawa Y, Fukutsu K, Aoki M, Itsukaichi H, Eguchi-Kasai K, Ohara H, 
Yatagai E, Kanai T and Ando K  2000
Inactivation of aerobic and hypoxic cells from three different cell lines by 
accelerated $^{3}$He, $^{12}$C and $^{20}$Ne-ion beams
{\it Radiat. Res.} {\bf 154} 485-96

%*\bibitem{GEANT4-Documents:2006} 
\item[] GEANT4-Documents 2006 {\it http://geant4.web.cern.ch/geant4/G4UsersDocuments/Overview/html/}

%*\bibitem{GEANT4-Webpage:2006} 
\item[] GEANT4-Webpage 2006 {\it http://geant4.web.cern.ch/geant4/}

%*\bibitem{Griesmayer:and:Auberger:2004}
\item[]  Griesmayer E and Auberger T  2004 
The status of MedAustron
{\it Radiother. Oncol.} {\bf 73} S202-5

%*\bibitem{Haberer:etal:2004}
\item[]  Haberer T, Debus J, Eickhoff H, Jakel O, Schulz-Ertner D and Weber U 2004
The Heidelberg ion therapy center
{\it Radiother. Oncol.} {\bf 73} S186-90

%*\bibitem{Hahn:and:Ricci:1966}
\item[]  Hahn R L and Ricci E 1966
Interactions of $^3$He Particles with $^9$Be, $^{12}$C, $^{16}$O and $^{19}$F 
{\it Phys. Rev.} {\bf 146} 650-9.

%*\bibitem{Heeg:etal:2004}
\item[]  Heeg P, Eickhoff H and Haberer T 2004
Conception of heavy ion beam therapy at Heidelberg University (HICAT)
{\it Z. Med. Phys.}  {\bf 14} 17-24

%*\bibitem{Hishikawa:etal:2004}   
\item[]  Hishikawa Y, Oda Y, Mayahara H, Kawaguchi A, Kagawa K, Murakami M and Abe M 2004
 Status of the clinical work at Hyogo
{\it Radiother. Oncol.} {\bf 73}  S38-40

%*\bibitem{Hishikawa:etal:2002}
\item[]  Hishikawa Y, Kagawa K, Murakami M, Sakai H, Akagi T and Abe M  2002
Usefulness of positron-emission tomographic images after proton therapy
{\it Int. J. Radiat. Oncol. Biol. Phys.} {\bf 53} 1388-91

%*\bibitem{Kempe:etal:2007}
\item[] Kempe J, Gudowska I and Brahme A 2007
Depth absorbed dose and LET distributions of therapeutic $^1$H, $^4$He, 
$^{7}$Li and $^{12}$C beams
{\it Med. Phys.} {\bf 34} 183-92    

%*\bibitem{Knust:and:Machulla:1983}
\item[] Knust E J and Machulla H-J 1983
High yield production of $^{18}$F in water target via the
$^{16}$O($^3$He,p)$^{18}$F reaction
{\it Int. J. Appl. Radiat. Isot.} {\bf 34} 1627-8

%*\bibitem{Levin:and:Hoffman:1999}
\item[]  Levin C S and Hoffman E J 1999
 Calculation of positron range and its effect on the fundamental limit
 of positron emission tomography system spatial resolution
{\it Phys. Med. Biol.} {\bf 44} 781-99
  
%*\bibitem{Nishio:etal:2005}
\item[]  Nishio T, Sato T, Kitamura H, Murakami K and Ogino T 2005
Distributions of $\beta^+$ decayed nuclei generated in the CH$_2$ and 
H$_2$O targets by the target nuclear fragment reaction using therapeutic 
MONO and SOBP proton beam
{\it Med. Phys.} {\bf 32} 1070-82

%*\bibitem{Oelfke:etal:1996}
\item[]  Oelfke U, Lam G K and Atkins M S  1996
Proton dose monitoring with PET: quantitative studies in Lucite
{\it Phys. Med. Biol.} {\bf 41} 177-96
  
%\bibitem{Osgood:etal:1964}
\item[]  Osgood D R, Patterson J R and Titterton E W 1964
The excitation function for reaction C$^{12}$(He$^3$,n$_0$)O$^{14}$
between threshold and 11.45 MeV
{\it Nucl. Phys.} {\bf 60} 503-8

%*\bibitem{Parodi:and:Enghardt:2000}
\item[] Parodi K and Enghardt W  2000
Potential application of PET in quality assurance of proton therapy
{\it Phys. Med. Biol.} {\bf 45} N151-6

%*\bibitem{Parodi:etal:2002}
\item[] Parodi K, Enghardt W and Haberer T  2002
In-beam PET measurements of $\beta^+$ radioactivity induced by proton beams
{\it Phys. Med. Biol.} {\bf 47} 21-36

%*\bibitem{Parodi:2004}
\item[] Parodi K 2004
On the feasibility of dose quantification with in-beam PET data in radiotherapy 
with $^{12}$C and proton
beams 2004, Ph.D. Dissertation, Technische Universit\"{a}t Dresden. 

%*\bibitem{Parodi:etal:2007a}
\item[] Parodi K, Ferrari A, Sommerer F and Paganetti H 2007a
Clinical CT-based calculations of dose and positron emitter distributions in proton therapy 
using the FLUKA Monte Carlo code 
{\it Phys. Med. Biol.} {\bf 52} 3369-87

%*\bibitem{Parodi:etal:2007b}
\item[]
Parodi K, Paganetti H, Shih H A, Michaud S, Loeffler J S, DeLaney T F, Liebsch N J, 
Munzenrider J E, Fischman A J, Knopf A and Bortfeld T 2007b
Patient study of in vivo verification of beam delivery and range, using positron emission 
tomography and computed tomography imaging after proton therapy
{\it Int. J. Radiat. Oncol. Biol. Phys.} {\bf 68} 920-34

%*\bibitem{Pawelke:etal:1996}
\item[] Pawelke J, Byars L, Enghardt W, Fromm W D, Geissel H, Hasch B G, Lauckner K, 
Manfrass P, Schardt D and Sobiella M  1996
The investigation of different cameras for in-beam PET imaging
{\it Phys. Med. Biol.} {\bf 41} 279-96

%*\bibitem{Pawelke:etal:1997}
\item[] Pawelke J, Enghardt W, Haberer T, Hasch B G, Hinz R, Kramer M, 
Lauckner K and Sobiella M   1997
In-beam PET imaging for the control of heavy-ion tumour therapy
{\it IEEE Trans. Nucl. Sci.} {\bf 44} 1492-8

%*\bibitem{Poenisch:etal:2004}
\item[] P\"onisch F, Parodi K, Hasch B G and Enghardt W 2004
The modelling of positron emitter production and PET imaging during carbon ion therapy 
{\it Phys. Med. Biol.} {\bf 49} 5217-32 
   
%*\bibitem{Pshenichnov:etal:2005}
\item[] Pshenichnov I, Mishustin I and Greiner W 2005
Neutrons from fragmentation of light nuclei in tissue-like media: a study with the GEANT4 toolkit
{\it Phys. Med. Biol.} {\bf 50} 5493-507

%*\bibitem{Pshenichnov:etal:2006}
\item[] Pshenichnov I, Mishustin I and Greiner W 2006
Distributions of positron-emitting nuclei in proton and carbon-ion therapy
studied with GEANT4
{\it Phys. Med. Biol.} {\bf 51} 6099-112

%*\bibitem{Schulz-Ertner:etal:2004} 
\item[] Schulz-Ertner D, Nikoghosyan A, Thilmann C, Haberer T, Jakel O, 
Karger C, Kraft G, Wannenmacher M and Debus J 2004
Results of carbon ion radiotherapy in 152 patients
{\it Int. J. Radiat. Oncol. Biol. Phys.} {\bf 58} 631-40

%*\bibitem{Shen:etal:1989} 
\item[] Shen W, Wang B, Feng J, Zhan W, Zhu Y and Feng E
Total reaction cross-section for heavy-ion collisions and its relation to the neutron 
excess degree of freedom
{\it Nucl. Phys.} A {\bf 491} 130-46 

%*\bibitem{Tomitani:etal:2003}
\item[] Tomitani T, Pawelke J, Kanazawa M, Yoshikawa K, Yoshida K, Sato M, 
Takami A, Koga M, Futami Y, Kitagawa A, Urakabe E, Suda M, Mizuno H, Kanai T, 
Matsuura H, Shinoda I and Takizawa S 2003
Washout studies of $^{11}$C in rabbit thigh muscle implanted by 
secondary beams of HIMAC
{\it Phys. Med. Biol.} {\bf 48} 875-89

%*\bibitem{Tripathi:etal:1997}
\item[] Tripathi R K, Cucinotta F A and Wilson J W 1999
Universal parameterization of absorption cross sections,
NASA Technical Paper 3621

%*\bibitem{Tsujii:etal:2004}
\item[]  Tsujii H, Mizoe J E, Kamada T, Baba M, Kato S, Kato H, Tsuji H, 
Yamada S, Yasuda S, Ohno T, Yanagi T, Hasegawa A, Sugawara T, Ezawa H, 
Kandatsu S, Yoshikawa K, Kishimoto R and Miyamoto T 2004
Overview of clinical experiences on carbon ion radiotherapy at NIRS.
{\it Radiother. Oncol.} {\bf 73} S41-9

%*\bibitem{Weisskopf:and:Ewing:1940} 
\item[] Weisskopf V E and Ewing D H  1940  On the yield of nuclear reactions with 
heavy elements {\it Phys. Rev.} {\bf 57} 472-85

%*\bibitem{Wellisch:and:Axen:1996} 
\item[] Wellisch H P and Axen D 1996 Total reaction cross section calculations in
 proton-nucleus scattering
{\it Phys. Rev.} C {\bf 54} 1329-32

%*\bibitem{Yashima:etal:2003} 
\item[] Yashima H, Uwamino Y, Iwase H, Sugita H, Nakamura T, Ito S and Fukumura A 2003
Measurement and calculation of radioactivities of spallation products 
by high-energy heavy ions
{\it Radiochim. Acta} {\bf 91} 689-96

%*\bibitem{Yashima:etal:2004} 
\item[] Yashima H, Uwamino Y, Iwase H, Sugita H, Nakamura T, Ito S, Fukumura A 2004
Cross sections for the production of residual nuclides by high-energy heavy ions
{\it Nucl. Instr. Methods} B {\bf 226} 243-63

\endrefs

\end{document}